\documentclass{JMLFS}

\journal{JMLFS-ID}
\vol{2020}

\received{xx January 2018}
\published{xx March 2018}

\def\be{\begin{equation}}
\def\ee{\end{equation}}
\def\bea{\begin{eqnarray}}
\def\eea{\end{eqnarray}}

\usepackage{amsmath} 
\usepackage{booktabs}  
\usepackage{subcaption, caption} 
\usepackage{tikz} 
\usetikzlibrary{shapes.geometric, arrows} 
\usepackage[ruled,vlined]{algorithm2e} 
\usepackage{hyperref} 
\hypersetup{
    colorlinks=true,
    linkcolor=blue,
    filecolor=blue,      
    urlcolor=blue,
    citecolor=blue,
    pdfpagemode=FullScreen,
    }

\usetikzlibrary{positioning} 
\tikzstyle{layer} = [rectangle, rounded corners, minimum width=3cm, minimum height=1cm,text centered, draw=black, fill=red!30]
\tikzstyle{physics} = [rectangle, rounded corners, minimum width=3cm, minimum height=1cm, text centered, draw=black, fill=blue!30]
\tikzstyle{output} = [rectangle, rounded corners, minimum width=3cm, minimum height=1cm, text centered, draw=black, fill=green!30]
\tikzstyle{arrow} = [thick,->,>=stealth]
\tikzstyle{process} = [rectangle, minimum width=3cm, minimum height=1cm, text centered, draw=black, fill=blue!30]
\tikzstyle{arrow1} = [thick,->,>=stealth]

\begin{document}

\title{Advancing Physics Data Analysis through Machine Learning and Physics-Informed Neural Networks}

\author{Vasileios Vatellis,\auno{1,2}}
\address{$^1$Institute of Informatics and Telecommunications,
  National Centre for Scientific Research “Demokritos”, \\15310 Aghia Paraskevi, Athens, Greece}
\address{$^2$School of Rural, Surveying and\\ Geoinformatics Engineering, National Technical University of Athens,\\ Zografou 157 73, Greece\\}

\begin{abstract}
In an era increasingly focused on green computing and explainable AI, revisiting traditional approaches in theoretical and phenomenological particle physics is paramount. This project evaluates various machine learning (ML) algorithms—including Nearest Neighbors, Decision Trees, Random Forest, AdaBoost, Naive Bayes, Quadratic Discriminant Analysis (QDA), and XGBoost—alongside standard neural networks and a novel Physics-Informed Neural Network (PINN) for physics data analysis. We apply these techniques to a binary classification task that distinguishes the experimental viability of simulated scenarios based on Higgs observables and essential parameters. Through this comprehensive analysis, we aim to showcase the capabilities and computational efficiency of each model in binary classification tasks, thereby contributing to the ongoing discourse on integrating ML and Deep Neural Networks (DNNs) into physics research. In this study, XGBoost emerged as the preferred choice among the evaluated machine learning algorithms for its speed and effectiveness, especially in the initial stages of computation with limited datasets. However, while standard Neural Networks and Physics-Informed Neural Networks (PINNs) demonstrated superior performance in terms of accuracy and adherence to physical laws, they require more computational time. These findings underscore the trade-offs between computational efficiency and model sophistication.
\end{abstract}

\maketitle

\begin{keyword}
Machine Learning\sep Artificial Intelligence\sep Computer Science
\end{keyword}

\section{Introduction}
Particle physics involves complex phenomena, such as neutrino masses and dark matter, which challenge existing theoretical frameworks. To explore these phenomena, it is essential to test various theoretical models against experimental data. This study focuses on enhancing the efficiency of this testing process by integrating machine learning (ML) techniques. Specifically, we introduce an ML filter to streamline the identification of experimentally viable parameter sets, improving computational efficiency and model validation in particle physics simulations.

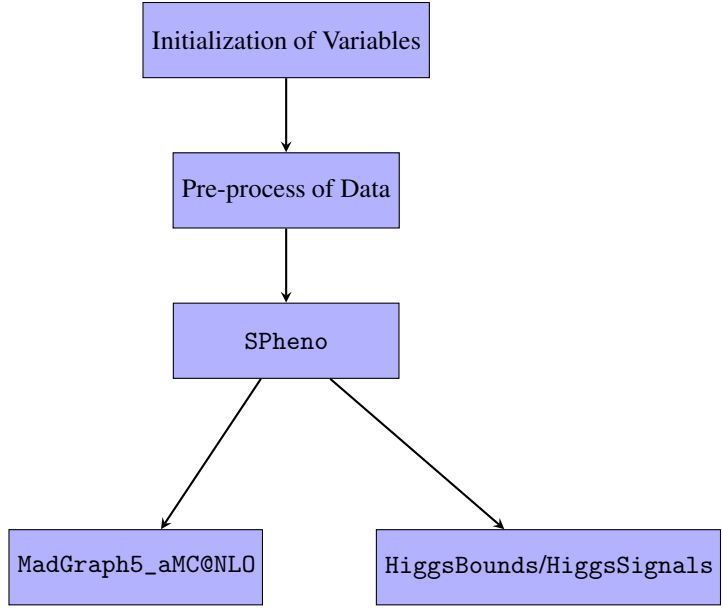
\begin{figure}
\centering
\begin{tikzpicture}[node distance=2cm]

\tikzstyle{process} = [rectangle, minimum width=3cm, minimum height=1cm, text centered, draw=black, fill=blue!30]
\tikzstyle{arrow} = [thick,->,>=stealth]
\node (init) [process] {Initialization of Variables};
\node (preprocess) [process, below of=init] {Pre-process of Data};
\node (spheno) [process, below of=preprocess] {\texttt{SPheno}};
\node (madgraph) [process, below=of spheno, xshift=-2cm] {\texttt{MadGraph5\_aMC@NLO}}; 
\node (higgs) [process, right=of madgraph, xshift=-0.5cm] {\texttt{HiggsBounds}/\texttt{HiggsSignals}}; 

\draw [arrow] (init) -- (preprocess);
\draw [arrow] (preprocess) -- (spheno);
\draw [arrow] (spheno) -- (madgraph);
\draw [arrow] (spheno) -- (higgs);

\end{tikzpicture}
\vspace{12pt} 
\caption{Process Flowchart}
\vspace{12pt}
\label{fig:process_flowchart}
\end{figure}

The initial stage of the physics analysis involves randomly initializing the free parameters of our theoretical model and computing the tree-level masses. These computed values are then input into the \texttt{SPheno}~\cite{spheno1,spheno2} code, which calculates decay modes for supersymmetric particles and Higgs bosons, as well as production cross-sections in electron-positron annihilation scenarios. \texttt{SPheno} also integrates with other tools, such as \texttt{HiggsBounds}~\cite{higgsBounds} and \texttt{MadGraph5\_aMC@NLO}~\cite{MadGraph}, to evaluate the compatibility of Higgs boson observations with theoretical predictions and simulate high-energy particle collisions. This integrated approach enables a comprehensive examination of the model, as shown in Fig.\ref{fig:process_flowchart}.
While the method offers high accuracy, it comes at a significant computational cost. To improve efficiency, we introduce a machine learning (ML) filter at the initialization stage, which preliminarily assesses whether the randomly initialized variables are likely to be experimentally viable. By applying this filter, we can focus computational resources on promising parameter sets, thus optimizing the overall process. Although the detailed physics computations are integral to our work, they are not the primary focus of this study. For a deeper understanding of the physics calculations, we refer readers to Ferreira et al.~\cite{ferreira2022phenomenology}. Our main emphasis is on leveraging machine learning to enhance the efficiency of identifying experimentally viable parameter sets.
\section{Dataset Description and Preparation}
The dataset employed in this study was derived from the comprehensive phenomenological analysis conducted by Ferreira et al. \cite{ferreira2022phenomenology,ferreira2023collider}, focusing on a Branco-Grimus-Lavoura (BGL) model. This model, a variant within the two Higgs doublet framework augmented by a single scalar singlet, stands out for its rich phenomenology in particle physics. The original dataset encompasses both experimental outcomes and the corresponding set of free parameters instrumental in computing these values.

For our analysis, we conducted an extensive data-cleaning process, in which we retained only a subset of variables relevant to our work from the initial dataset. This refinement involved selecting the essential free parameters and specific physics values (denoted by $\mathbf{P}$) relevant for predictive modeling. The selected points were tested based on electroweak precision observables (or oblique parameters), Higgs observables, and quark flavor violation (QFV) constraints included in $\mathbf{P}$, categorizing each instance as either experimentally viable or nonviable according to predefined experimental criteria. This preprocessing step was crucial to ensure that our models focused on the free parameters, thereby enhancing the accuracy and interpretability of the analysis within the framework of the BGL model. Notably, our dataset contains approximately 16\% viable points, indicating an imbalance. We represent the dataset as $\mathbf{X}$, where $\mathbf{X} = {x_1, \dots, x_N}$, and $N$ is the number of simulated data points. A supplementary dataset contains the physics observables, represented as $\mathbf{P}$, such that for each data point $x_i$ with $i \in {1, \dots, N}$, there is an associated set of physics observables $\mathbf{p}_i = {p^i_1, p^i_2, p^i_3}$, with $i$ ranging from 1 to $N$. Further details regarding the restrictions and the physics can be found in Ferreira et al.~\cite{ferreira2022phenomenology}.
\section{Methodology}
As part of our study, we conducted a comprehensive evaluation of various machine learning algorithms to determine their effectiveness in predicting outcomes based on our dataset. The assessment process began with the preparation of the dataset, followed by the application of standard scaling to normalize the feature space. This preprocessing step is crucial for models that are sensitive to feature magnitude. We implemented a stratified k-fold cross-validation approach with five folds for evaluating the ML algorithms. This method ensures a robust assessment while maintaining the original dataset's class proportion within each fold, thus addressing potential biases in model performance due to imbalanced class distributions. Our selection of ML models ranged from simpler algorithms, such as Gaussian Naive Bayes and Decision Trees, to more complex ensemble methods including Random Forest, AdaBoost, and XGBoost~\cite{XGBoost}. Each model was rigorously tested to gain a comprehensive understanding of its performance, focusing on metrics such as accuracy, precision, recall, F1 score, and the area under the ROC curve (ROC AUC). These metrics were selected for their capacity to provide a detailed insight into model performance, particularly in the classification "valid point" or "no-valid point." tasks where the inherent imbalance in our dataset needs to be addressed, as balancing true positive and false positive rates is essential. Creating synthetic data would introduce bias, and therefore, it was not employed in our study. We systematically recorded the evaluation results, noting not only the mean and standard deviation of each metric across the folds but also the time required for each model's training and evaluation. This method allowed us to make nuanced comparisons between the models, considering both their predictive performance and computational efficiency.The hyperparameters used for training each of these machine learning models are summarized in Table~\ref{tab:hyperparameters}.

For the deep neural networks segment of our study, we adopted a focused approach to assess the training procedure, primarily utilizing the validation loss as a key indicator of performance. This method involves dividing our dataset into training, validation, and test subsets, with the latter serving to further validate our results and as a proxy to estimate the model's generalization ability on unseen data. Throughout the training epochs, we closely monitored the validation loss, aiming to minimize it as an objective measure of the model’s prediction error. A distinctive emphasis was placed on exploring the capabilities of Physics-Informed Neural Networks (PINNs)~\cite{pinnRaissi} alongside traditional Neural Networks (NNs) to address the unique challenges presented by our physics-based dataset. In subsection~\ref{sec:PINN}, we delve into the structural details of these models and their operational dynamics.

\begin{figure}
\captionsetup{} 
\centering
\begin{subfigure}{.185\textwidth}
    \caption{}
  \begin{tikzpicture}[node distance=2cm]
    \node (input) [output] {Input Layer};
    \node (hidden1) [layer, below of=input] {Hidden Layer 1};
    \node (hidden2) [layer, below of=hidden1] {Hidden Layer 2};
    \node (hidden3) [layer, below of=hidden2] {Hidden Layer 3};
    \node (output) [output, below of=hidden3] {Output Layer};
    \draw [arrow] (input) -- (hidden1);
    \draw [arrow] (hidden1) -- (hidden2);
    \draw [arrow] (hidden2) -- (hidden3);
    \draw [arrow] (hidden3) -- (output);
  \end{tikzpicture}
        \label{fig:nn}
\end{subfigure}
\begin{subfigure}{.21\textwidth}
    \caption{}
  \begin{tikzpicture}[node distance=2cm]
    \node (input) [output] {Input Layer};
    \node (hidden1) [layer, below of=input] {Hidden Layer 1};
    \node (hidden2) [layer, below of=hidden1] {Hidden Layer 2};
    \node (physics) [physics, below of=hidden2] {Physics-Informed Layer};
    \node (output) [output, below of=physics] {Output Layer};
    \draw [arrow] (input) -- (hidden1);
    \draw [arrow] (hidden1) -- (hidden2);
    \draw [arrow] (hidden2) -- (physics);
    \draw [arrow] (physics) -- (output);
  \end{tikzpicture}
        \label{fig:pinn}
\end{subfigure}
\caption{Comparison of a standard Neural Network Fig~\ref{fig:nn} and a Physics-Informed Neural Network (PINN) Fig~\ref{fig:pinn}.}
\label{fig:NN_vs_PINN}
\end{figure}
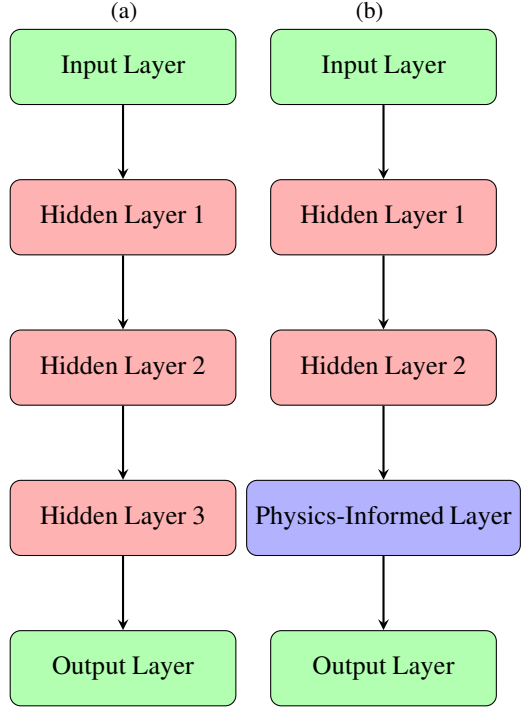
\subsection{Physics-Informed Neural Networks (PINNs)}
\label{sec:PINN}
The cornerstone of Physics-Informed Neural Networks (PINNs) lies in their innovative integration of physical laws and information directly into the learning process. This integration is meticulously encapsulated within our dual-output model architecture. This architecture, visualized in Fig~\ref{fig:NN_vs_PINN}, encapsulates the essence of PINNs and illustrates the differences with traditional neural networks (NNs). The PINN enables the simultaneous prediction of physics-informed outputs $\hat{\mathbf{P}}$ alongside the primary task-specific predictions $\hat{\mathbf{Y}}$, ensuring compliance with underlying physical principles:
\begin{eqnarray}
\begin{split}
\hat{\mathbf{P}} & = \mathcal{G}_w(\mathbf{X}) \Leftrightarrow \hat{\mathbf{p}}_i = \mathcal{G}_w(x_i), \\
\hat{\mathbf{Y}} & = \mathcal{F}_w(\mathbf{X}, \hat{\mathbf{P}}) \Leftrightarrow \hat{y}_i = \mathcal{F}_w(x_i, \hat{\mathbf{p}}_i).
\end{split}
\end{eqnarray}
The critical innovation in our PINN approach (Fig~\ref{fig:nodesPINN}) occurs during the training phase (Alg~\ref{Alg:PINN}), particularly in the structuring of the loss function. In addition to the conventional loss term for measuring the accuracy of primary task predictions (e.g., classification accuracy or mean squared error for regression), an additional term is introduced to ensure that the predictions made by the physics-informed layer—P1, P2, P3 as shown in Fig~\ref{fig:nodesPINN}—closely match the actual physics quantities derived from the data or known physical laws. To effectively balance the contribution of this physics-based loss term with the primary task loss, we include a normalization constant, $\lambda$. In this study, $\lambda$ is set to one, simplifying the integration process of these terms and maintaining a direct influence of the physics-informed constraints on the overall loss function.
\begin{eqnarray}\label{eq:vcg}
\begin{split}
\mathcal{L}_{\text{total}} &= \mathcal{L}_{\text{classification}} + \lambda \mathcal{L}_{\text{physics}},\\
 &= \mathcal{L}_{\text{classification}}(\mathbf{Y},\hat{\mathbf{Y}}) + \lambda \mathcal{L}_{\text{physics}}(\mathbf{P},\hat{\mathbf{P}}), \\
 & = \frac{1}{N}\sum_i^N{\mathcal{L}_{\text{classification}}(y_i,\mathcal{F}_w(x_i,\mathcal{G}_w(x_i))) + \lambda \mathcal{L}_{\text{physics}}(\mathbf{p}_i,\mathcal{G}_w(x_i))}.
\end{split}
\end{eqnarray}
Here, $\mathcal{L}_{\text{classification}}$ is given by the Binary Cross-Entropy with Logits Loss function, while $\mathcal{L}_{\text{physics}}$ is the mean-squared-error between the outputs of the physics-informed layer and the physical variables as computed in Ferreira et al. \cite{ferreira2022phenomenology}. By incorporating this additional physics-based loss term, the PINN is explicitly trained not only to achieve accuracy in its primary tasks but also to approximate physical values and make decisions informed by physics. This approach enhances both the interpretability and reliability of the model’s outputs, bridging the gap between complex data analysis and fundamental physical principles.

The training process diverges from the evaluation process, by focusing exclusively on the classification loss during the evaluation phase. This strategic omission of the physics loss term during model evaluation allows for a direct and clear comparison between the performance metrics of standard Neural Networks (NNs) and Physics-Informed Neural Networks (PINNs), ensuring our evaluations are objective and insightful.

\begin{figure}
\centering
\captionsetup{belowskip=12pt}
\begin{tikzpicture}[node distance=1.5cm]
     \node (input) [draw, circle] {Input};
    
    \node (hidden1-1) [draw, circle, below of=input, xshift=-2cm] {H1-1};
    \node (hidden1-2) [draw, circle, right of=hidden1-1, xshift=0.5cm] {H1-2};
    \node [right of=hidden1-2, xshift=-0.5cm] {...};
    \node (hidden1-3) [draw, circle, right of=hidden1-2, xshift=1cm] {H1-N};
    
    \node (hidden2-1) [draw, circle, below of=hidden1-1] {H2-1};
    \node (hidden2-2) [draw, circle, right of=hidden2-1, xshift=0.5cm] {H2-2};
    \node [right of=hidden2-2, xshift=-0.5cm] {...};
    \node (hidden2-3) [draw, circle, right of=hidden2-2, xshift=1cm] {H2-N};
    
    \node (physics1) [draw, circle, below of=hidden2-1, fill=red!30, xshift=0.5cm] {P1};
    \node (physics2) [draw, circle, right of=physics1, fill=red!30] {P2};
    \node (physics3) [draw, circle, right of=physics2, fill=red!30] {P3};
    
    \node (output) [draw, circle, below of=physics2] {Output};
    
    \draw[->] (input) -- (hidden1-1);
    \draw[->] (input) -- (hidden1-2);
    \draw[->] (input) -- (hidden1-3);
    \draw[->] (hidden1-1) -- (hidden2-1);
    \draw[->] (hidden1-2) -- (hidden2-2);
    \draw[->] (hidden1-3) -- (hidden2-3);
    \draw[->] (hidden2-1) -- (physics1);
    \draw[->] (hidden2-2) -- (physics2);
    \draw[->] (hidden2-3) -- (physics3);
    \draw[->] (physics1) -- (output);
    \draw[->] (physics2) -- (output);
    \draw[->] (physics3) -- (output);
\end{tikzpicture}
\caption{Architecture of the Physics-Informed Neural Network (PINN). The network comprises an input layer, followed by multiple fully connected hidden layers (H1 and H2), and a specialized physics-informed layer (highlighted in red), concluding with an output layer. While the illustration shows selective connections for clarity, it is important to note that each layer is fully connected to its subsequent layer. The physics-informed layer (P1, P2, P3) is designed to encode specific physical principles or constraints relevant to the problem domain, facilitating the integration of domain knowledge directly into the learning process.}
    \label{fig:nodesPINN}
\end{figure}
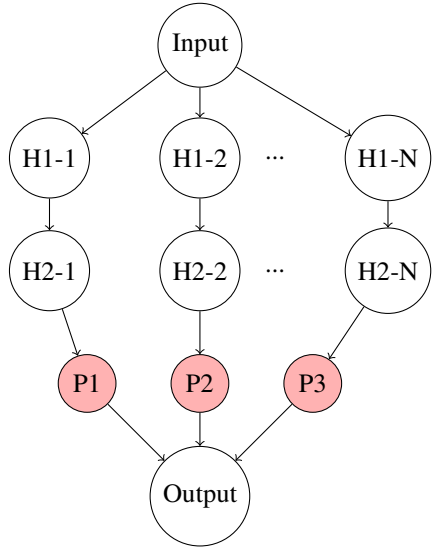
\begin{algorithm}
\caption{Training a Physics-Informed Neural Network (PINN)}
\label{Alg:PINN}
\KwData{Training data $\mathbf{X}=\{x_1,\dots,x_N\}$, Training labels $\mathbf{Y}=\{y_1,\dots,y_N\}$, Physics data $\mathbf{P}=\{\mathbf{p}_1,\dots,\mathbf{p}_N\}$}
\textbf{Training:} Tuning of $w$ based on $\hat{y}_i= f_w(x_i, p_i)$ and $\hat{\mathbf{p}}_i = \mathcal{G}_w(x_i)$\;
// Start of training step\\
\textbf{for all}{ $x_i\ \in \ X$ with $i \ \in \{1,\dots,N\}$ \textbf{do}\\
    \hspace{0.5cm} Forward pass to compute predictions $\hat{y}_i$ and physics-informed outputs $\hat{p}_i$\;
    \hspace{0.5cm} Compute classification loss $\mathcal{L}_{\text{classification}}(y_i, \hat{y}_i) = \mathcal{L}_{\text{classification}}(y_i, f_w(x_i,\hat{p}_i))$\;
    \hspace{0.5cm} Compute physics loss $\mathcal{L}_{\text{physics}}(p_i, \hat{\mathbf{p}}_i)= \mathcal{L}_{\text{physics}}(p_i,\mathcal{G}_w(x_i))$\;
    \hspace{0.5cm} Compute total loss $\mathcal{L}_{\text{total}} = \mathcal{L}_{\text{classification}} + \lambda \mathcal{L}_{\text{physics}}$\;
    \hspace{0.5cm} Backpropagation\;
\textbf{end}
}
\end{algorithm}

\section{Results and Discussion}
\begin{table}
\centering
\begin{tabular}{lc}
\hline
\textbf{Model} & \textbf{Hyperparameters} \\ \hline
Nearest Neighbors & n\_neighbors = 3 \\ 
Decision Tree & max\_depth = 5 \\ 
Random Forest & max\_depth = 5, n\_estimators = 10, max\_features = 1 \\ 
AdaBoost & \text{Default hyperparameters} \\ 
Naive Bayes & \text{Default hyperparameters} \\ 
QDA & \text{Default hyperparameters} \\ 
XGBoost & use\_label\_encoder = False, eval\_metric = 'logloss' \\ \hline
\end{tabular}
\vspace{8pt}
\caption{Hyperparameters used for training machine learning models of Table~\ref{tab:ML_results}.}
\label{tab:hyperparameters}
\end{table}
\begin{table}
\centering
\resizebox{\textwidth}{!}{
\begin{tabular}{lcccccc}
\hline
\textbf{Model} & \textbf{Accuracy} & \textbf{Precision} & \textbf{Recall} & \textbf{F1 Score} & \textbf{ROC AUC} & \textbf{Time (seconds)} \\ \hline
Nearest Neighbors & 0.8359 $\pm$ 0.0026 & 0.5160 $\pm$ 0.0091 & 0.4461 $\pm$ 0.0074 & 0.4785 $\pm$ 0.0033 & 0.7924 $\pm$ 0.0066 & 629.70 \\
Decision Tree & 0.8781 $\pm$ 0.0007 & 0.6851 $\pm$ 0.0039 & 0.5140 $\pm$ 0.0060 & 0.5874 $\pm$ 0.0037 & 0.9015 $\pm$ 0.0032 & 149.38 \\
Random Forest & 0.8333 $\pm$ 0.0063 & 0.3493 $\pm$ 0.8596 & 0.0130 $\pm$ 0.0401 & 0.0249 $\pm$ 0.0761 & 0.8463 $\pm$ 0.0090 & 27.13 \\
AdaBoost & 0.8576 $\pm$ 0.0026 & 0.6488 $\pm$ 0.0243 & 0.3409 $\pm$ 0.0270 & 0.4466 $\pm$ 0.0202 & 0.8804 $\pm$ 0.0021 & 580.41 \\
Naive Bayes & 0.6059 $\pm$ 0.0045 & 0.2953 $\pm$ 0.0024 & \textbf{0.9636 $\pm$ 0.0042} & 0.4521 $\pm$ 0.0029 & 0.8174 $\pm$ 0.0048 & \textbf{2.01} \\
QDA & 0.6183 $\pm$ 0.0058 & 0.3012 $\pm$ 0.0034 & 0.9562 $\pm$ 0.0100 & 0.4581 $\pm$ 0.0045 & 0.8173 $\pm$ 0.0034 & 8.01 \\
XGBoost & \textbf{0.9608 $\pm$ 0.0014} & \textbf{0.8990 $\pm$ 0.0121} & 0.8651 $\pm$ 0.0105 & \textbf{0.8817 $\pm$ 0.0037} & \textbf{0.9882 $\pm$ 0.0009} & 9.24 \\ \hline
\end{tabular}
}
\vspace{8pt}
\caption{Performance Metrics of Various Machine Learning Models in Physics Data Analysis. This table presents a comparative overview of the accuracy, precision, recall, F1 score, ROC AUC, and computation time for each model. Top-performing values in each category are emphasized in bold.}
\label{tab:ML_results}
\end{table}
In our evaluation of machine learning models for physics data analysis which can be seen in Table~\ref{tab:ML_results}, we focused on critical metrics such as Precision, Recall, and F1 Score, which are essential for accurately identifying true positive scenarios. XGBoost excelled among the models tested, achieving the highest F1 Score (0.8817 ± 0.0037). This score demonstrates an optimal balance between precision and recall, pivotal for minimizing classification errors in our study. Moreover, XGBoost also led with the best ROC AUC score (0.9882 ± 0.0009), underscoring its exceptional ability to differentiate between classes across various threshold settings. It additionally recorded the highest precision (0.8990 ± 0.0121). While the Naive Bayes model displayed the highest recall, it did so at the expense of precision, highlighting the trade-offs involved in model selection. Overall, XGBoost proved to be the most effective model for our specific analytical requirements, offering high reliability, robust discrimination capabilities, and an outstanding balance of performance metrics, requiring also 9.24 sec for training. 
\begin{table}
\centering
\begin{tabular}{c c c c c c }
\toprule
\textbf{Learning Rate} & \textbf{Accuracy} & \textbf{Precision} & \textbf{Recall} & \textbf{F1 Score} & \textbf{ROC AUC} \\ 
\midrule
0.0001 & \textbf{0.9740} & 0.9404 & 0.9040 & \textbf{0.9219} & \textbf{0.9938} \\
0.0002 & 0.9736 & 0.9531 & 0.8879 & 0.9193 & 0.9936 \\
0.0005 & 0.9716 & 0.9217 & \textbf{0.9099} & 0.9158 & 0.9937 \\
8e-05 & 0.9682 & \textbf{0.9549} & 0.8530 & 0.9011 & 0.9924 \\
\bottomrule
\end{tabular}
\vspace{8pt}
\caption{Performance Metrics of Top Neural Network Models Under Different Learning Rates. This table showcases the results for neural network configurations that demonstrated the best overall performance in our analysis. Values in bold represent the highest scores achieved in each respective performance category across all tested models, indicating the most effective model settings for each metric."}
\label{tab:NN_Performance}
\end{table}

Our analysis of the neural network models which can be seen in Table~\ref{tab:NN_Performance}, examined under various learning rates, reveal small yet crucial differences in performance metrics that impact their efficacy in scenario classification. The model with a learning rate of 0.0001 outperforms others, achieving the highest F1 Score (0.9219) and Precision (0.9404), alongside an impressive Recall (0.9040), indicating its strong capability to balance between identifying true positives and minimizing false outcomes. Its ROC AUC of 0.9938 also demonstrates superior discriminatory power.

In our study, while XGBoost offers a balanced performance across all metrics, it is outpaced by almost all the neural network models tested. Notably, the neural network with a learning rate of 0.0001 not only demonstrated superior performance with the highest F1 Score and Precision but also achieved an impressive Recall (0.9040), indicating its exceptional capability in balancing true positive identification and minimizing false outcomes. Its ROC AUC of 0.9938 further underscores its superior discriminatory power over XGBoost. Although these neural networks generally require more training time than XGBoost, their superior performance across almost all metrics makes them highly effective for applications where model accuracy is prioritized over training speed.
\begin{table}[t]
\centering
\begin{tabular}{c c c c c c c c}
\toprule
\textbf{LR} & \textbf{Act. Func.} & \textbf{Phys. Act. Func.} & \textbf{Accuracy} & \textbf{Precision} & \textbf{Recall} & \textbf{F1 Score} & \textbf{ROC AUC} \\ 
\hline
0.0009 & GELU & GELU & \textbf{0.9711} & \textbf{0.9145} & 0.9152 & \textbf{0.9148} & \textbf{0.9933} \\
0.004 & GELU & GELU & 0.9684 & 0.8853 & \textbf{0.9351} & 0.9095 & 0.9924 \\
0.005 & ReLU & LeakyReLU & 0.9694 & 0.9178 & 0.9003 & 0.9089 & 0.9894 \\
0.004 & ReLU & LeakyReLU & 0.9706 & 0.9013 & 0.9286 & 0.9147 & 0.9924 \\
0.006 & ReLU & ELU & 0.9619 & 0.8546 & 0.9345 & 0.8928 & 0.9913 \\
0.003 & ReLU & ELU & 0.9608 & 0.8705 & 0.9032 & 0.8866 & 0.9910 \\
0.005 & GELU & GELU & 0.9630 & 0.8605 & 0.9331 & 0.8953 & 0.9904 \\
0.003 & GELU & GELU & 0.9658 & 0.8746 & 0.9319 & 0.9024 & 0.9926 \\
0.002 & GELU & GELU & 0.9675 & 0.8894 & 0.9235 & 0.9061 & 0.9899 \\
0.006 & ReLU & LeakyReLU & 0.9676 & 0.9172 & 0.8894 & 0.9031 & 0.9911 \\
\bottomrule
\end{tabular}
\vspace{8pt}
\caption{Performance Comparison of PINN Models with Varying Activation Functions. This table displays the accuracy, precision, recall, F1 scores, and ROC AUC for Physics-Informed Neural Networks (PINNs) across different learning rates and activation function configurations. The best values in each performance category are highlighted in bold, emphasizing the models' top performances in our evaluation.}
\label{tab:summary_pinn_models}
\end{table}

In our expanded analysis of Physics-Informed Neural Network (PINN) models detailed in Table~\ref{tab:summary_pinn_models}, we assessed various configurations to gauge how learning rates and activation functions impact essential metrics like Precision, Recall, and the F1 Score. The top PINN model, with a learning rate of 0.0009 and GELU as both the activation and physics activation function, stands out by achieving the highest F1 Score (0.9148). This score reflects an optimal balance, with equally high Precision (0.9145) and Recall (0.9152), indicating effective integration of physical principles into the learning process. The superior performance of GELU can be attributed to its Gaussian Distribution-Based Smoothing, which weights inputs through a probabilistic lens, enhancing the model's ability to handle the complex, non-linear relationships inherent in physics data. Additionally, GELU supports smoother gradient flow during backpropagation compared to other activation functions. This characteristic prevents issues related to vanishing or exploding gradients, facilitating more stable and effective learning across deeper network architectures. These properties make GELU particularly suitable for PINNs, where capturing subtle physical behaviors accurately is crucial.

Comparatively, this top PINN model outperforms the best XGBoost configuration, which had an F1 Score of 0.8817, demonstrating superior balance and precision-recall trade-off. When matched against the top-performing neural network model, which reached an F1 Score of 0.9219, the leading PINN model shows competitive precision but a slightly lower overall F1 score. Despite this, the unique strength of PINNs in incorporating physical laws directly into their framework provides an explainability advantage against both XGBoost and top-performing neural network.

These results highlight the effectiveness of PINNs in optimizing both sensitivity and specificity, crucial for reducing false negatives and positives in complex systems and the ability to make decisions based on physics information without losing effectiveness. Such capabilities make PINNs particularly valuable in physics data analysis, where ensuring high accuracy in prediction under physical laws is paramount.

In summary, for the initial stages of computation, where only a small portion of the dataset is available, XGBoost is highly recommended as the machine learning filter. Its speed and relative accuracy make it ideal for quick, preliminary filter. On the other hand, for long-term simulations where comprehensive datasets are collected, Physics-Informed Neural Networks (PINNs) are advisable. PINNs leverage the underlying physical laws to enhance the predictive capability of traditional neural networks, integrating crucial physics-based insights into the otherwise non-transparent ML models. This approach ensures that the models are not only accurate but also fulfil the established physical principles, making them particularly valuable for in-depth, physics-intensive investigations.

\section{Future Work}
Looking ahead, the exploration of several intriguing research directions promises to enrich the domain of physics data analysis further. One of the most compelling prospects involves investigating the transferability of the top-performing neural network and PINN models to other physical systems with similar Lagrangians. This endeavour would not only highlight the generalizability and adaptability of these models but also deepen our comprehension of their fundamental workings. Moreover, determining the minimal dataset size necessary to achieve approximately 90\% accuracy emerges as a fascinating challenge, especially in the case of XGBoost, which we recommended as a preliminary filter. This inquiry would entail a comparison between the time savings from reduced simulation requirements and the computational demands of model training, aiming to optimize data usage efficiency. Additionally, the integration of uncertainty quantification within PINNs could significantly enhance their predictive power by offering probabilistic bounds on predictions, thereby bolstering confidence in model outputs and facilitating more informed decision-making. The adoption of ensemble methods that fuse the strengths of diverse models, including PINNs and traditional machine learning algorithms, may also unveil synergistic benefits, potentially elevating overall model performance and robustness. Lastly, applying PINNs in applications, spanning from particle physics experiments to cosmological data analysis, would furnish a comprehensive testing ground for these advanced computational tools, pushing the frontiers of current technology and knowledge in a concerted effort to unravel the complexities of our universe.

\bibliographystyle{unsrt}  
\bibliography{references}

\end{document}